\documentclass[showpacs,twocolumn,showkeys,amsmath,amssymb,pra,superscriptaddress,nofootinbib]{revtex4} 
\usepackage{amsmath}
\usepackage{color}
\usepackage{graphicx}   
\usepackage{dcolumn}    
\usepackage{bm}         
\usepackage{changes}

\begin{document} 
 
\title[]{Effects of interactions on the generalized Hong-Ou-Mandel effect}
\author{B. Gertjerenken}
\affiliation{Department of Mathematics and Statistics, University of Massachusetts Amherst, Amherst, Massachusetts 01003-9305, USA}
\affiliation{Institut f\"ur Physik, Carl von Ossietzky Universit\"at, D-26111 Oldenburg, Germany}
\author{P.G. Kevrekidis}
\affiliation{Department of Mathematics and Statistics, University of Massachusetts Amherst, Amherst, Massachusetts 01003-9305, USA}
\affiliation{Center for Nonlinear Studies and Theoretical Division, Los Alamos
National Laboratory, Los Alamos, NM 87544}

                  
\date{\today}
 
\begin{abstract}
We numerically investigate the influence of interactions on the generalized Hong-Ou-Mandel (HOM) effect for bosonic particles and show results for the cases 
of $N=2$, $N=3$ and $N=4$ bosons interacting with a beam splitter, whose role is played
by a $\delta$-barrier. In particular, we focus on the effect of attractive interactions and compare the results with the repulsive case, as well as with the 
analytically available results for the non-interacting case (that we use as a 
benchmark). We observe a fermionization effect both for growing repulsive and attractive interactions, i.e., the dip in the HOM coincidence count is 
progressively smeared out, for increasing interaction strengths.
The role of input asymmetries is  also explored.
\end{abstract}


\maketitle


\section{Introduction}
When two indistinguishable photons simultaneously hit a 50~\% - 50~\% 
beam splitter, one in each input port, they interfere destructively and no coincidence counts of particles detected in both output ports can be observed. This inherently quantum mechanical effect is a striking manifestation of the bosonic properties of the photons. Since its first observation in 1987, the 
celebrated Hong-Ou-Mandel (HOM)~\cite{HongOuMandel87} effect has triggered a multitude of investigations and has been the focus of ongoing research: it has found application in the creation of post-selected entanglement between photon pairs~\cite{KwiatEtAl95}, and can be exploited for logic gates in linear optical quantum computation~\cite{KnillEtAl01}, or to demonstrate the purity of a solid-state single-photon source~\cite{SantoriEtAl02}. The HOM effect has also been experimentally demonstrated with single photons emitted by two independently trapped single atoms~\cite{BeugnonEtAl06}. The influence of varying properties of the beam-splitter on the joint probability distribution (of finding both photons on the same side) has been investigated in~\cite{LongoEtAl12}. Experimental realizations have been extended to larger particle numbers, namely three photons impinging on a multiport mixer~\cite{Campos2000}, and to one and two-photon pairs~\cite{CosmeEtAl08}. A review on multi-photon experiments and the generalized Hong-Ou-Mandel effect is given in~\cite{Ou2007}.

More recently, the HOM effect has been generalized to massive particles, i.~e. to $N$ bosons or fermions passing simultaneously through a symmetric Bell multiport beam splitter~\cite{LimEtAl05}, and to a large number of particles impinging on a single beam splitter~\cite{LaloeEtAl12,LaloeEtAl12b}. In the latter case it has been shown for the balanced beam-splitter that if an even number of particles impinges on the beam-splitter from either side, an even number must also emerge from each side. A recent proposal suggests to observe the HOM effect with colliding Bose-Einstein condensates~\cite{LewisSwanEtAl14} for a set-up that has already been used to demonstrate the violation of the Cauchy-Schwarz inequality~\cite{KheruntsyanEtAl12}.

Effects of the interparticle interaction have been investigated in~\cite{CompagnoBanchiBose2014} within a Bose-Hubbard set-up where a transition from bunching to antibunching can be observed for growing repulsive interparticle interactions, corresponding to a fermionization of the particles~\cite{CominottiEtAl14,BuschHuyetEtAl03}. Bosonic atoms have also been proposed as a suitable candidate for Knill-Laflamme-Milburn quantum computation, 
with the advantage of very controllable input state preparation~\cite{Popescu07}. It is relevant to highlight here that two-particle quantum interference
akin to the Hong-Ou-Mandel effect has recently been observed
in the context of independently prepared bosonic atoms
in tunnel-coupled optical tweezers~\cite{kaufmanetal14}. The latter
experiment follows up on another experimental realization of
such quantum interference effects in the context of 
electrons~\cite{bocquillonetal13}. It should thus be clear
that  the remarkable advances in
the setting of ultracold gases enable the very accurate
counting~\cite{markus13} and controllability~\cite{nature12,greiner09} of bosonic
atoms towards such experiments. Motivated by the fact that the parity of the number of atoms in a potential well is experimentally accessible 
more straightforwardly than the exact particle number it has been pointed out that parity measurements can yield useful signatures of the generalized HOM effect~\cite{LaloeEtAl12b}. Recent experimental results even demonstrate the simultaneous determination of the number of atoms in each well of a double-well trap with single-atom resolution for up to $N=500$ atoms per well~\cite{StroescuEtAl15}.

Here, we will chiefly focus on the effect of \emph{attractive} interparticle interactions on the HOM dip and demonstrate that also in this case a transition from bunching to antibunching with growing interaction strength can be observed. This can again be interpreted as a fermionization effect, in agreement with~\cite{TempfliEtAl08}, where the Bose-Fermi mapping has been demonstrated for attractive 1D bosons. The latter generalization enables the formation of 
gas-like states that fermionize as the attraction becomes stronger.
It is that, somewhat counter-intuitive (on the basis of the
nature of the interaction) feature that we also 
find in the present setting. Our investigation focuses on the (generalized) 
HOM effect for a 1D Bose gas on the $N$-particle level, enabling true quantum 
behavior. We focus on the cases $N=2$,~$N=3$ and~$N=4$. In addition
to examining the somewhat less studied attractive case, we compare
the results with those of the repulsive case and importantly
with the non-interacting case that can be addressed, in principle,
in its full generality for arbitrary $N$, as will be discussed below. 

In the original HOM experiment the photons emerge as a result of 
quantum interference in a superposition of states $|2,0\rangle$ and $|0,2\rangle$ and thus in a measurement would always be found on the same side. 
As an aside, we note in passing, effectively regarding
the large $N$ limit of mean-field matter waves that
a recent proposal suggests an analogue of the HOM effect with bright solitons~\cite{SunEtAl14}. In this classical case it is found that the indistinguishability of the particles yields a 0.5 split mass on either side for 
solitary waves (by parity symmetry at this mean-field level). 
But for very slight deviations it can be observed that all the particles 
always end up on the same side of the barrier potential. 
Recently, also the collisional dynamics of matter-wave solitons (in the absence of a barrier potential) have been investigated experimentally~\cite{NguyenEtAl14}. Additionally, 
the setting of interactions of individual single~\cite{cornish}
and multi-component~\cite{engels} solitary waves with barriers (that play
the role of the beam splitter here) is certainly within experimental
reach.
Collisions have also been suggested as a way to create entanglement between indistinguishable solitons~\cite{LewensteinMalomed09} and initially independent and indistinguishable boson pairs~\cite{Holdaway13}. Collisions of distinguishable solitons have been proposed as a possibility to create mesoscopic Bell states~\cite{GertjerenkenEtAl13}.

Our presentation is structured as follows: Sec.~\ref{sec:model} gives an introduction to the model system, the Lieb-Liniger(-McGuire) model, and a quantum beam-splitter. In Sec.~\ref{sec:results} we present numerical results for the collision of two monomers, two dimers and the asymmetric case of a monomer and a dimer on an additional barrier potential. Sec.~\ref{sec:conclusion} 
summarizes our findings and presents a number of future challenges.

\section{Model system}\label{sec:model}
The $N$-particle dynamics of interacting bosons in (quasi-)one-dimensional geometries (corresponding to a 3D geometry with tightly confined radial degrees of freedom) can be described within the exactly solvable Lieb-Liniger(-McGuire) model~\cite{LiebLiniger63,McGuire64}
\begin{eqnarray}
\hat{H}_{\rm{LL}} &=& -\sum_{j=1}^{N} \frac{\hbar^2}{2m}\partial^2_{x_j} + \sum_{j=1}^{N-1}\sum_{n=j+1}^{N}g\delta(x_j-x_n). \nonumber \\
\end{eqnarray}
Here, contact interaction between the $N$ bosons is assumed and quantified with the interaction parameter~$g$. In the following we will investigate the scattering dynamics within an additional harmonic confinement 
(emulating the typical parabolic trap relevant to experimental
settings; cf.~\cite{cornish,engels}) at a repulsive delta-like barrier potential in the middle of the harmonic confinement, yielding the full Hamiltonian
\begin{equation}\label{eq:LLwithpot}
\hat{H} = \hat{H}_{\rm{LL}} + \sum_{j=1}^{N}V_{\rm{ext}}(x_j)
\end{equation}
with the external potential $V_{\rm{ext}}(x_j) = \frac{1}{2}m\omega^2x_j^2+v_0\delta(x_j)$.

For the numerical implementation the system can be discretized via the Bose-Hubbard Hamiltonian 
\begin{eqnarray}\label{eq:BHH}
\hat{H}_{\rm{discretized}} & = & -J \sum_j(\hat{c}^{\dagger}_j \hat{c}^{\phantom{\dagger}}_{j+1} + \hat{c}^{\dagger}_{j+1} \hat{c}^{\phantom{\dagger}}_{j}  ) + \frac{U}{2}\sum_j\hat{n}_j(\hat{n}_j-1) \nonumber \\&&+A\sum_j \hat{n}_j j^2 + v_0\delta_{j,0},
\end{eqnarray} 
where $\hat{c}^{(\dagger)}_j$ denotes the annihilation (creation) operator for lattice site $j$, $\hat{n}_j = \hat{c}^{\dagger}_j\hat{c}_j$ is the particle number operator, and $U$ the on-site interaction strength. The tunneling strength is given by $J \sim \hbar^2/2mb^2$ with lattice spacing~$b$ for $b\rightarrow 0$, $A\equiv\frac{1}{2}m\omega^2 b^2$ defines the strength of the harmonic confinement and $v_0$ the strength of the delta-like barrier potential. For sufficiently small lattice spacing $b\rightarrow0$ the Lieb-Liniger model~(\ref{eq:LLwithpot}) with additional harmonic confinement is recovered.

\subsection{Quantum beam-splitter}\label{sec:beamsplitter}
\begin{figure}
\begin{center}
\includegraphics[width = 0.7\linewidth]{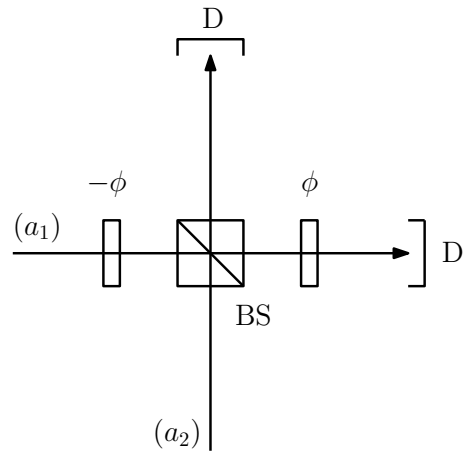}
\end{center}
\caption{Schematic depiction of a quantum beam-splitter (BS) with input modes $(a_1)$ and $(a_2)$, detectors (D) at both outputs and optional additional phase shifts.}    
\label{fig:beamsplitter}
\end{figure}

The repulsive delta-like barrier potential acts as a beam-splitter. In this section we present a theoretical description for the noninteracting case, following~\cite{HarocheRaimond06} (see also~\cite{LaloeEtAl12,LaloeEtAl12b}). Assume a generalized beam-splitter with two incoming modes $(a_1)$ and $(a_2)$ being coupled at the beam-splitting device and possible additional phase shifts, as depicted in Fig.~\ref{fig:beamsplitter}. This system is described by the time-evolution operator
\begin{equation}
 U_c(\theta,\phi) = \begin{pmatrix}
  \cos(\theta/2) & i\mathrm{e}^{i\phi}\sin(\theta/2) \\
  i\mathrm{e}^{-i\phi}\sin(\theta/2)  & \cos(\theta/2)
 \end{pmatrix},
\end{equation}
where $\theta$ is related to the complex transmission and reflection coefficients~$t=\cos(\theta/2)$ and $r=i\sin(\theta/2)$, obeying $|t|^2+|r|^2=1$ and $tr^{\ast}+t^{\ast}r = 0$.
It will be seen that the occupation probabilities of the possible output states are independent of the phase~$\phi$.

We first review the well-known case of a balanced beam-splitter with one particle in each input mode. The output state then reads
\begin{eqnarray}
&&U_c(\pi/2,0)|1,1\rangle \\ & = & \frac{1}{\sqrt{2}}\left(i\left(a_1^{\dagger}\right)^2 +a_1^{\dagger}a_2^{\dagger} -a_2^{\dagger} a_1^{\dagger} +i\left(a_2^{\dagger}\right)^2  \right)|0,0\rangle
\end{eqnarray}
For the bosonic case the application of the corresponding commutation relations, $[a^{\phantom{\dagger}}_i,a^{\dagger}_j]\equiv a^{\phantom{\dagger}}_ia^{\dagger}_j-a^{\dagger}_ja^{\phantom{\dagger}}_i=\delta_{ij}$ and $[a^{\dagger}_i,a^{\dagger}_j]=[a^{\phantom{\dagger}}_i,a^{\phantom{\dagger}}_j]=0$, yields the charateristic bunching:
\begin{equation}\label{eq:N2bosons}
U_c(\pi/2,0)|1,1\rangle  = \frac{i}{\sqrt{2}}\left(|2,0\rangle + |0,2\rangle\right).
\end{equation}
Both particles would always be detected in the same output mode. For fermions obeying the commutation relations $\{a^{\phantom{\dagger}}_i,a^{\dagger}_j\}\equiv a^{\phantom{\dagger}}_ia^{\dagger}_j+a^{\dagger}_ja^{\phantom{\dagger}}_i=\delta_{ij}$ and $\{a^{\dagger}_i,a^{\dagger}_j\}=\{a^{\phantom{\dagger}}_i,a^{\phantom{\dagger}}_j\}=0$, the output state exhibits anti-bunching:
\begin{equation}\label{eq:N2fermions}
U_c(\pi/2,0)|1,1\rangle  = -|1,1\rangle.
\end{equation}
For bosons this calculation can be straightforwardly extended to larger particle numbers: consider a Fock input state with $n$ particles in mode~$(a_1)$ and $m$ particles in mode~$(a_2)$. The output state then is given by the superposition state
\begin{eqnarray}\label{eq:outputstate}
&&U_c(\theta,\phi)|n,m\rangle\nonumber \\
&&= \sum_{l=0}^n\sum_{k=0}^m\binom{n}{l}^{1/2}\binom{m}{k}^{1/2}\binom{k+l}{k}^{1/2}\binom{m-k+n-l}{m-k}^{1/2}\nonumber \\
&&\times (i\sin(\theta/2))^{m-k+l}\cos(\theta/2)^{k+n-l}\exp(i\phi(m-k-l))\nonumber \\
&&\times |m-k+n-l,k+l\rangle.
\end{eqnarray}
This corresponds to the occupation probability
\begin{eqnarray}\label{eq:prob_analytical}
 && p_{|n_f,m_f\rangle}\nonumber \\
&=&|\langle n_f,m_f|U_c(\theta,\phi)|n,m\rangle|^2\nonumber \\
&=& n!m!n_f!m_f!\cos(\theta/2)^{2(m_f+n)}\sin(\theta/2)^{2(m-m_f)}\nonumber \\
&&\times\left[\sideset{}{'}\sum^n_{l=0}\frac{(-1)^l\sin(\theta/2)^{2l}\cos(\theta/2)^{-2l}}{l!(n-l)!(m_f-l)!(m-m_f+l)!}  \right]^2,
\end{eqnarray}
of a final state $|n_f,m_f\rangle$, in agreement with Eq.~(14) from~\cite{LaloeEtAl12} for the balanced beam-splitter. A derivation of the occupation probabilities for arbitrary reflectivity/transmittivity has also been given in~\cite{LaloeEtAl12b}. The prime indicates that summands resulting in undefined negative factorials do not contribute.

For indistinguishable fermions on the other hand an extension of this description to $N>2$ is not directly possible for this two-mode set-up due to the Pauli principle: an occupation of any mode with more than one particle is not allowed. Note that also for the description of $N$ fermions in a multi-mode setting as~\cite{LimEtAl05} the number of input / output modes has to be equal to or larger than the number of particles.

\begin{figure}
\begin{center}
\includegraphics[width = 1.0\linewidth]{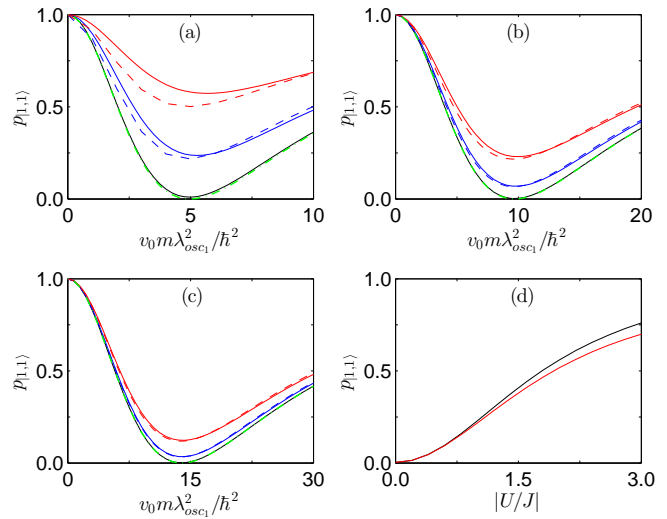}
\end{center}
\caption{(Color online) Analogue of Hong-Ou-Mandel effect for two monomers colliding on a narrow delta-like potential barrier situated in the middle of an additional harmonic confinement with $\omega=1.0$. The monomers are initially displaced by equal amounts $|x_0|$ to the left and right of the barrier potential. Shown is the occupation probability $p_{\rm{|1,1\rangle}}$ of state $|1,1\rangle$ at $t/T = 0.5$ vs. the scaled height $v_0m\lambda^2_{osc_1}/\hbar^2$ of the barrier potential for interaction strengths $U/J = 0.0$ (solid black line), $U/J = -0.5$ (solid blue line) and $U/J = -1.0$ (solid red line) and different initial displacements (a)~$|x_0|/\lambda_{\rm{osc,1}} = 5$, (b)~$|x_0|/\lambda_{\rm{osc,1}} = 10$ and (c)~$|x_0|/\lambda_{\rm{osc,1}} = 15$ of the monomers. Results for repulsive interactions $U/J = 0.5$ (dashed blue line) and $U/J = 1.0$ (dashed red line) are displayed as well. Dashed green line: analytical prediction given by Eq.~(\ref{eq:prob_analytical}). (d) Occupation probability of state $|1,1\rangle$ vs. ratio $|U/J|$ of interaction strength to tunneling strength for attractive (black line) and repulsive interactions (red line) for initial displacements~$|x_0|/\lambda_{\rm{osc,1}} = 10$.}
\label{fig:hom1}
\end{figure}
\begin{figure}
\begin{center}
\includegraphics[width = 1.0\linewidth]{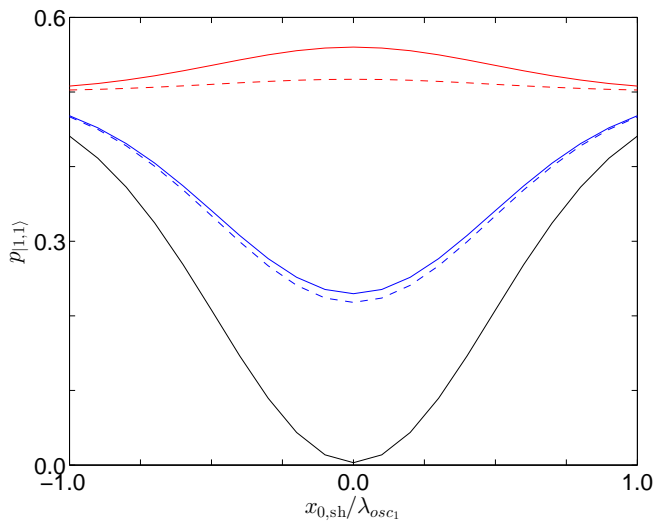}
\end{center}
\caption{(Color online) Influence of shifts $x_{0,\rm{sh}}/\lambda_{osc_1}$ around the initial position $x_0/\lambda_{osc_1}=10$ of the right particle for monomers on the occupation probability of state $|1,1\rangle$. The height $v_0m\lambda^2_{osc_1}/\hbar^2=200$ is chosen to ensure $50\%-50\%$-splitting in the case of no shift. Black: $U/J = 0.0$, blue: $U/J = -1.0$ (solid line), $U/J = 1.0$ (dashed line), red: $U/J = -2.0$ (solid line), $U/J = 2.0$ (dashed line). Same parameters as in Fig.~\ref{fig:hom1}~(b).}    
\label{fig:hom1b}
\end{figure}

\begin{figure}
\begin{center}
\includegraphics[width = 1.0\linewidth]{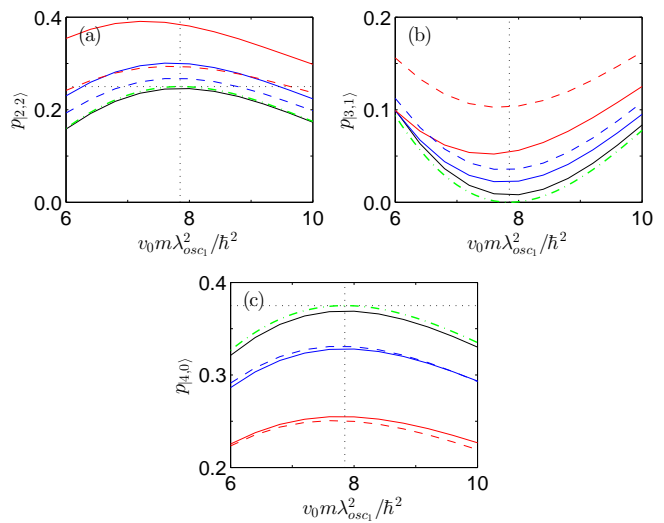}
\end{center}
\caption{(Color online) Same as Fig.~\ref{fig:hom1} for two dimers and $\omega = 4.0$. Initial displacement by $2.2\lambda_{osc_1}$. Displayed is the occupation probability of the states $|2,2\rangle$, $|3,1\rangle$ and $|4,0\rangle$ at $t/T = 0.5$ vs. the scaled height $v_0m\lambda^2_{osc_1}/\hbar^2$ of the barrier potential for interactions strengths $U/J = 0.0$ (black), $U/J = -0.5$ (blue), $U/J = -1.0$ (red), $U/J = 0.5$ (dashed blue) and $U/J = 1.0$ (dashed red). Vertical dotted black line: Value of potential height for 50\%-50\%-splitting. Horizontal dotted black line: Analytical expectation for the linear case and a balanced beam splitter. Note that $p_{\rm{|3,1\rangle}} = p_{\rm{|3,1\rangle}} = 0$ is the analytical prediction for this case. Dashed green line: analytical prediction given by Eq.~(\ref{eq:prob_analytical}). The slight deviation from the analytical prediction is a remaining lattice effect due to the restriction of the numerics for $N=4$ to a smaller number of lattice sites, here 75.}    
\label{fig:hom4}
\end{figure}
\section{Numerical results}\label{sec:results}
Our numerical results are obtained by implementing~(\ref{eq:BHH}) for sufficiently small lattice spacing~$b$, to emulate the Lieb-Liniger model. 
First, we investigate the direct analogue to the original HOM experiment~\cite{HongOuMandel87}. We simulate the collision of two single particles on a barrier potential situated in the middle~$x=0$ of an additional harmonic confinement. The initial state is given as the (symmetrized) product state of two single particles, each of them initially prepared in well-separated harmonic confinements with same trapping frequency and potential minima located at $x = \pm x_0$. Subsequently, the harmonic confinements are replaced by a single harmonic confinement with potential minimum at $x=0$ implying that both particles then are located at the classical turning points of the harmonic trap and start to accelerate with oscillation period
$T=\frac{2\pi}{\omega}$
towards the trap center where they simultaneously hit the narrow, delta-like potential barrier. In Fig.~\ref{fig:hom1} we display the occupation probability of state~$|1,1\rangle$ at $t=T/2$ for different interaction and potential
strengths and different initial displacements resulting in different kinetic energies when the particles hit the barrier at $t=T/4$. Without interparticle interactions we observe a behavior analogous to the original Hong-Ou-Mandel experiment: if the potential height is chosen to ensure 50\%-50\%-splitting for each single particle, we clearly observe the HOM dip with a negligible occupation probability of state~$|1,1\rangle$; the system is found in the superposition state~(\ref{eq:N2bosons}), nicely illustrating the bosonic symmetry and interference properties of the particles, in the absence of interactions. Again analogous to the original HOM effect changes in potential height, resulting in deviations from 50\%-50\%-splitting, lead to a growing contribution of~$|1,1\rangle$, in agreement with the analytical prediction~(\ref{eq:prob_analytical}).
Notice that both for this and for all the cases that will follow, 
the difference between the numerically obtained solid black line
and the analytically derived green dash-dotted line on the basis
of Eq.~(\ref{eq:prob_analytical}) will be a measure of the quality of
our approximation (via the Bose-Hubbard underlying lattice model) of the 
continuum setting. In all the case examples displayed below,
these deviations are very small.

For repulsive interactions a transition from bunching to antibunching has been predicted with growing interactions~\cite{CompagnoBanchiBose2014}. 
A particularly intriguing feature of our findings is that {\it also} in 
the presence of attractive interactions the Hong-Ou-Mandel dip is suppressed and a transition from bunching to antibunching can be observed. 
Such a somewhat counter-intuitive (on the basis
of the attractive interactions) phenomenon is reminiscent of 
similar fermionization features 
as those reported in~\cite{TempfliEtAl08}; cf. Figs. like Fig. 1b therein.
For comparison, 
Fig.~\ref{fig:hom1} shows results both for attractive and for repulsive interactions. It can be seen that for increasing kinetic energy of the particles the suppression of the HOM dip is reduced, cf. Fig.~\ref{fig:hom1}~(b) and Fig.~\ref{fig:hom1}~(c). 
This is natural to expect since in the latter setting, the kinetic effects
dominate the interaction ones and progressively (as the kinetic
energy increases) restore the non-interacting picture.
While differences between repulsive and attractive interactions are found for lower kinetic energies attractive and repulsive interactions yield quantitatively similar results for larger kinetic energies. This can be explained by the shorter interaction time in the middle of the harmonic trap for larger kinetic energies, rendering interactions less influential. Summarizing these 
results, the HOM dip is suppressed with growing attractive interaction strengths for all cases, but this suppression is most effective 
for low center-of-mass kinetic energies. Fig.~\ref{fig:hom1}~(d) compares the suppression of the HOM dip for continously growing attractive and repulsive interaction strengths. It can be seen that up to $|U/J|\approx 1$ the occupation probability of state~$|1,1\rangle$ is the same but for larger interaction strengths the HOM dip is suppressed equally or in some cases slightly more
strongly in the presence of attractive interactions (in comparison
to repulsive ones).  For repulsive interactions the $|1,1\rangle$ configuration should be energetically favored, while for attractive interactions the superposition state~(\ref{eq:N2bosons}) should intuitively 
be favored. We attribute our observation of the antibunching to the fact that, 
in analogy to the Bose-Fermi mapping in the 
repulsive case~\cite{CominottiEtAl14,BuschHuyetEtAl03} (for large
interactions $U$), also attractive bosons in 1D undergo a fermionization 
for growing interaction strengths~\cite{TempfliEtAl08}.

At the mean-field level, applicable for large $N$, 
a 0.5 split mass on either side is the result for a balanced beam-splitter (due to parity symmetry). Yet, it has recently 
been shown that for very slight deviations from indistinguishability,
it can be observed that almost all the particles may end up on the same side of the barrier potential, constituting a mean-field analogue of the 
HOM-effect~\cite{SunEtAl14}. This observation prompted us to 
investigate the role of asymmetries in the set-up by introducing slightly 
different initial displacements of the monomers, cf.~Fig.~\ref{fig:hom1b}: due to the harmonic confinement the particles then still hit the barrier at the same point of time but with different kinetic energies. While on the mean-field level such asymmetries have caused the occurrence of HOM-like behavior, here we observe a complementary behavior: at the few-particle quantum-mechanical
level such asymmetries result in a suppression of the HOM dip. For very large 
interaction strengths~$|U/J|\approx 2$ the occupation  $p_{\rm{|1,1\rangle}}$ 
remains close to its already large value of 0.51 (repulsive) and 
0.55 (attractive) without shifts of the initial position. This suggests that as $N$ is increased a transition should arise
between the small-$N$ behavior presenting a dip at vanishing asymmetry for $N=2$ and odd-even oscillations for $N > 2$~\cite{LaloeEtAl12}
and the large $N$ behavior bearing a $0.5$ splitting fraction at vanishing
asymmetry and a potentially close to unity fraction for slight asymmetries.
Unfortunately, this regime is not currently accessible to our techniques,
yet would constitute an intriguing theoretical (and experimental)
problem to consider in future work.

\begin{figure}
\begin{center}
\includegraphics[width = 1.0\linewidth]{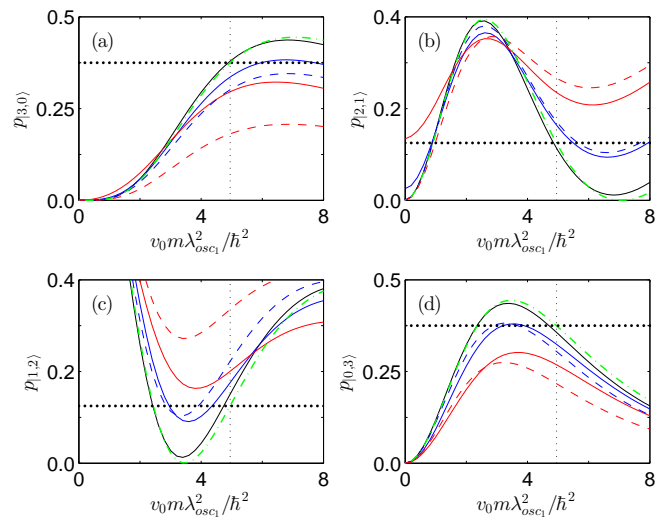}
\end{center}
\caption{(Color online) Same as Figs.~\ref{fig:hom1} and \ref{fig:hom4} and for a dimer (initially left) and a monomer (initially right) and  $\omega = 1.5$. Initial displacement by $3.3\lambda_{osc_1}$. Displayed is the occupation probability of the states $|3,0\rangle$, $|2,1\rangle$, $|1,2\rangle$ and $|0,3\rangle$ at $t/T = 0.5$ vs.~the scaled height $v_0m\lambda^2_{osc_1}/\hbar^2$ of the barrier potential for interactions strengths $U/J = 0.0$ (black), $U/J = -0.5$ (blue), $U/J = -1.0$ (red), $U/J = 0.5$ (dashed blue) and $U/J = 1.0$ (dashed red). Calculated on 121 lattice points. Vertical dotted black line: Value of potential height for 50\%-50\%-splitting. Horizontal dotted black line: Analytical expectation for the linear case and a balanced beam splitter. Dashed green line: analytical prediction given by Eq.~(\ref{eq:prob_analytical}). The slight deviation from the analytical prediction again is a residual lattice effect.}    
\label{fig:hom3}
\end{figure}

We now turn to the investigation of the situation of two dimers, cf. Fig.~\ref{fig:hom4}. In this case, the ground state of each of the dimers is obtained by imaginary time evolution: the many-particle initial wave function is taken as a product wave function of both dimer ground states. While for the monomers a large number of lattice sites was accessible, for the dimers the numerically accessible size of the Hilbert space is considerably reduced. We choose 75 lattice sites. Once again, we utilize the non-interacting case as a benchmark
for our results, finding good agreement with the analytical prediction
of Eq.~(\ref{eq:prob_analytical}). The slight deviation is caused by 
residual lattice effects. Due to symmetry we only display occupation probabilities of the states~$|2,2\rangle$,~$|3,1\rangle$ and $|4,0\rangle$. It can be seen that for the balanced beam-splitter only states with even particle numbers on either side of the beam-splitter contribute to the output state. This corresponds to the predicted odd-even oscillations for the generalized HOM effect in~\cite{LaloeEtAl12}. Fig.~\ref{fig:hom4} also displays the effect of interactions. It can be observed that the Fock-states are affected in a different manner: while the occupation probability of state~$|4,0\rangle$ (and $|0,4\rangle$) is affected in a similar way, regardless of the sign of the interaction, considerable differences can be observed for the other states. Importantly, we 
note that in this case no direct analog of the 
Hong-Ou-Mandel effect exists, as even in the absence of
interactions and for a 50~\% - 50~\% beam-splitter, the 
$|2,2\rangle$ state is not fully suppressed (although the 
asymmetric $|3,1\rangle$ and $|1,3\rangle$ ones are, as indicated above). 
What our
results suggest is that as we depart from this limit,
the preferred non-interacting state of $|4,0\rangle$
gets suppressed in favor of increased probabilities of both
the $|2,2\rangle$ and $|3,1\rangle$ states. The attractive case progressively leads to fermionization in the
form of the $|2,2\rangle$ state, while repulsive interactions
progressively lead to an increased probability of
 the $|3,1\rangle$  state.

As discussed in Sec.~\ref{sec:beamsplitter} the analytical description in terms of the two-mode quantum beam-splitter for fermions cannot be extended to cases with $N>2$. While this prevents a direct comparison of our results with the fermionic case, we nonetheless presume that the effects of interaction we observe can be attributed to an {\it effective} fermionization: Ref.~\cite{TempfliEtAl08} demonstrates the occurrence of fermionization for an attractive one-dimensional Bose gases by the formulation of a Bose-Fermi mapping and discusses the stationary states of a few bosons in a harmonic confinement.

Finally, we investigate the asymmetric case of $N=3$, the collision of a dimer and a monomer, cf.~Fig.~\ref{fig:hom3}, where again we interpret the result of interactions as suggesting an effective fermionization, following a similar reasoning as above. As the results for the interacting case are governed by a directional effect, we display the occupation probabilities of all possible final states. The non-interacting case again is in accordance with the analytical prediction~(\ref{eq:prob_analytical}). In the absence of a barrier, here, we get solely
the $|1,2\rangle$ state, while under a 50~\% - 50~\% splitting
(in the non-interacting case)  $|3,0 \rangle$ and
$|0,3\rangle$ are equally populated and favored in comparison
to $|1,2\rangle$ and $|2,1\rangle$ (who are also equally probably
between them).  The interactions generally favor less the
bunched states $|3,0 \rangle$ and
$|0,3\rangle$, although the different interaction signs lead to
a more pronounced effect in the former in comparison to the latter.
Similarly, the interactions lead to an increased probability of
$|1,2\rangle$ and $|2,1\rangle$, although again the difference
in signs affects more the former in comparison to the latter.

\section{Conclusion}\label{sec:conclusion}
We have numerically investigated the generalized HOM-effect for bosonic atoms in a (quasi-)one-dimensional geometry with a special emphasis on the influence of interactions. In accordance to existing literature for a repulsive Bose-Hubbard set-up~\cite{CompagnoBanchiBose2014}, we observe a transition from bunching to antibunching for repulsive interactions. In addition, we have presented numerical results that show such a fermionization and generally similar
anti-bunching trends also for attractive interactions. 
Our results have been illustrated beyond the well-known case
of $N=2$ atoms, for those with $N=3$ and $N=4$. The role of the
strength of the interactions (and its relative influence in comparison
to the kinetic energy) was illustrated and the influence of a potential
asymmetry in the initial configuration was also discussed. We believe
that this presentation, thus, yields a fairly complete picture of
the generalized form of HOM-type (numerical) experiments for small
atom numbers $N$.

The high controllability available in current experiments with atoms,
as discussed in the Introduction,
 appears to render very plausible and accessible 
an experimental observation of the generalized Hong-Ou-Mandel effect,  
exploring, e.g., odd-even oscillations in the occupation of the final state
in the non-interacting limit, as well as suppression of HOM-like
dips for increasing interaction strengths, as discussed herein. 
The control of interaction strengths via tools such as 
magnetically~\cite{inouye98}
or optically~\cite{fatemi00} induced Feshbach resonances should enable
the systematic observation of the numerically obtained features.
We argue that for larger displacements out of the trap center the interactions play a negligible role and results close to the linear case can be obtained.  Due to their stability, bright solitons are a particularly promising candidate to observe the odd-even oscillations in the regime of low particle numbers where 
quantum effects are important. Additionally, a systematic exploration 
of regimes of progressively larger $N$, to explore the systematics
of the transition from the small $N$ to the mean-field behavior would
be of particular interest for future studies.

\begin{acknowledgments}
After completing our work we discovered that related calculations were done by W. J. Mullin and F. Lalo\"{e}. We thank them for sharing their preprint with us. We thank R. Bisset for discussions. B. G. acknowledges support from the European Union through FP7-PEOPLE-2013-IRSES Grant Number 605096.
P.G.K. acknowledges support from the National Science Foundation
under grant DMS-1312856, from the European Union 
through FP7-PEOPLE-2013-IRSES Grant Number 605096, and from
the Binational
(US-Israel) Science Foundation through grant 2010239.
P.G.K.'s work at Los Alamos is supported in part by the
U.S. Department of Energy. The computations were performed on the HPC cluster HERO, located at the 
University of Oldenburg and funded by the DFG through its Major Research 
Instrumentation Programme (INST 184/108-1 FUGG), and by the Ministry of 
Science and Culture (MWK) of the Lower Saxony State.

\end{acknowledgments}

\bibliographystyle{unsrt}

\end{document}